\documentclass{appolb}
\usepackage{graphicx}
\usepackage{wasysym}
\usepackage{amssymb}

\begin{document}
\title{On the Possibility of Having Roman Pots\\around IP8 in Run 4 and Beyond%
\thanks{Presented at XXVI Cracow Epiphany Conference.}%
}
\author{Maciej Trzebi\'nski
\address{Institute of Nuclear Physics Polish Academy of Science\\ul. Radzikowskiego 152, 31-342 Krak\'ow, Poland.}
}
\maketitle
\begin{abstract}
Proton trajectories along LHC beam 1 (clockwise direction) in the vicinity of the LHCb Interaction Point (IP8) for the most recent LHC Run 4 optics (HLLHC1.5) were presented. On this basis, three possible locations of forward proton detectors were identified: 150, 180 and 430 m from IP8. For these locations geometric acceptances were estimated. For the proton relative energy loss, $\xi$, these limits are $0.05 < \xi \lesssim 0.1$, $0.025 < \xi \lesssim 0.1$ and $0.003 < \xi < 0.013$, respectively. The influence of boost of the central system due to the acceptance of LHCb detector on $\xi$ was discussed. Finally, the impact of pile-up was presented.

\end{abstract}
  
\section{Motivation}
Studies of diffractive and exclusive processes are an important part of the physics programme at hadron colliders. This is also true for the LHC, where such processes are being investigated by all major experiments.

Diffraction is usually connected to the exchange of a colorless object: photon in case of electromagnetic and a so-called Pomeron in case of strong interactions. A colorless exchange may lead to one of the most characteristic features of diffraction -- presence of a rapidity gap, defined as a space in rapidity devoid of any particles. In addition, since colorless exchange does not change the quantum numbers, the nature of interacting objects does not change as well. For example, if a colorless object is exchanged by colliding protons, they may stay intact, lose part of their energy and be scattered at very small angles (typically into the accelerator beam pipe).

Exclusive production is a process in which all particles can be measured. For proton-proton collisions, it is a reaction $p+p \to p + X + p$, where $X$ is a fully specified system of particles well separated in rapidity from the outgoing protons.

There are two major techniques used to recognize diffractive events: rapidity gap and forward proton tagging. The first one does not require installation of additional detectors. However, a gap may be destroyed by the particles coming from pile-up -- independent proton-proton interactions happening during the same bunch crossing. Moreover, a gap can be produced outside the acceptance of the central detector. The second technique -- direct proton measurement -- addresses to some extent the issues of the first one. For example, it can be efficiently applied in a non-zero pile-up environment. However, since protons are scattered at very small angles (about few hundreds microradians), there is a need to install additional equipment -- so-called forward proton detectors.

There are two major designs of forward proton detectors: Roman pots and Hamburg beampipes. The first one resembles a cylindrical pot. It is designed to be used in a so-called ``warm LHC sector'' -- a straight section, up to about 260 m from IP. The Hamburg beampipe is more suitable for the ``cold region'' (arc, beyond 260 m).

Roman pots are already present at the LHC. In ATLAS \cite{ATLAS} two systems of such detectors were installed: ALFA \cite{ALFA1, ALFA2} and AFP \cite{AFP}. At the LHC interaction point 5 (IP5), Roman pots are used by CMS \cite{CMS} and TOTEM \cite{TOTEM1, TOTEM2} groups. Hamburg beampipes were not yet in operation at the LHC. However, some initial LHC designs were done by \textit{e.g.} FP420 Collaboration \cite{FP420}.

Diffractive data were collected during LHC Run 1 and Run 2 and more is foreseen to come during Run 3. There are also concepts of keeping Roman pots around IP1 and IP5 at High Luminosity LHC (HL-LHC), \textit{i.e.} in Run 4 and beyond. However their presence is not yet decided, as foreseen harsh pile-up conditions ($\mu$ around 200) significantly limit the physics programme.

On the other hand, pile-up at the LHCb collision point should increase only from about 1 to about 5 in Run 4 \cite{LHCb_pileup}. Taking into account an excellent coverage of LHCb in the forward direction ($2 < \eta < 5$) \cite{LHCb}, having proton detectors installed in vicinity of LHCb should provide interesting diffractive data, complementary to the ones collected by ATLAS and CMS/TOTEM.

This publication intends to be a starting point for a discussion of having forward proton detectors at LHCb. In the next sections the properties of scattered protons will be discussed. On this basis, three possible detector locations will be described.

\section{Proton Trajectories}
The first step is to check the possible locations of the forward proton detectors. Vicinity of collision point is not empty. On contrary, there are plenty of elements: magnets, collimators, beam monitors, \textit{etc.} installed. In this paper, the newest publicly available layout was used: HL-LHC V1.5 \cite{lhc_optics}. It should be noted that the final HL-LHC layout is still being discussed \cite{hllhc_layout}. Following studies were performed for the centre-of-mass energy of $\sqrt{s} = 14$ TeV and the value of betatron function $\beta^* = 3$~m (see Ref. \cite{Trzebinski_optics} for explanations). Proton trajectories were calculated using the MAD-X program \cite{MADX} and the Polymorphic Tracking Code tool \cite{PTC}.

LHC beams circulate in two horizontally displaced beam pipes which join into a common one about 130 m away from the Interaction Point (IP). The beam performing the clockwise motion (viewed from above the ring) is called Beam 1 and the other one Beam 2. Sector between IP and $\sim 260$ m is called a \textit{warm region} or a \textit{straight section}. Elements located further than 260 m are in so-called \textit{cold region} or \textit{arc}.

The first magnets after IP, the focusing triplet, are positioned at about 40 m. Other quadrupoles (Q4, Q5 and Q6) are installed around 140 m, 170 m and 240 m from the LHCb IP (IP8). Two dipole magnets used for the beam separation are installed at about 70 m (D1) and 130 m (D2). From about 260~m a cold section starts. It contains sequences of dipoles and quadrupoles.

\begin{figure}[htb]
\centerline{%
  \includegraphics[width=\textwidth]{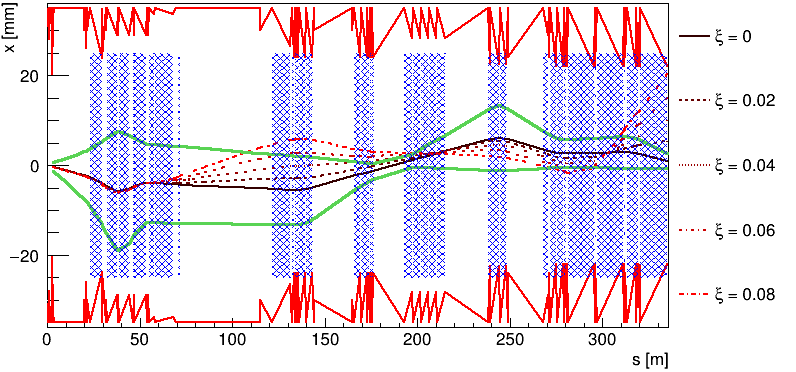}}
\caption{Energy dependence of the proton trajectory in $(x, s)$ plane for beam 1 around IP8 $(0,0)$. Protons were generated with the transverse momentum set accordingly to -115 $\mu$rad value of the crossing angle. For detailed description of the elements see text.}
\label{fig_prot_traj_1}
\end{figure}

Proton trajectories are usually described in a curvilinear, right handed coordinate system $(x, y, s)$. The $s$-axis points toward the beam direction, $x$ axis is in the bent plane and $y$ is perpendicular to the bent plane. As was shown in \textit{e.g.} Ref. \cite{Trzebinski_optics}, for the nominal collision optics the key factor is a deflection for a proton with an energy loss $\xi = 1 - E_{proton}/E_{beam}$. Beam~1 proton trajectories in vicinity of IP8 are shown in Fig. \ref{fig_prot_traj_1}. In this figure IP8 is located at $(0, 0)$, the LHC elements listed in the optics file \cite{lhc_optics} are marked as a hatched blue areas and the beampipe aperture is represented by the red lines. Nominal proton trajectory is drawn as a thick, black line. Its initial deflection from $x=0$ is due to a crossing angle of -115 $\mu$rad, which translates to the initial scattered proton momentum $p_x = -0.805$ GeV. Trajectories of protons which lost 2, 4, 6 and 8\% of their initial energy are drawn as a red dashed or dotted lines. Finally, thick green lines represent the 15$\sigma$ envelope, where $\sigma$ is nominal width of the beam (see Ref. \cite{Trzebinski_optics}).

The outgoing protons are scattered into the beam pipe on the inner side of the ring. Contrary to IP1 and IP5, this should give much easier access to the detector infrastructure.

As can be concluded from Fig. \ref{fig_prot_traj_1}, the possible placement of Roman pots can be, in principle, anywhere outside the hatched blue areas. Another limit is due to the minimal detector-beam distance. In this paper a conservative value of 15$\sigma$ was considered. For the simplicity one could assume that the forward proton detectors should be installed in places where diffractive trajectories are outside the 15$\sigma$ ``envelope'', \textit{e.g.} around 150 and 180 m.

\begin{figure}[htb]
\centerline{%
  \includegraphics[width=\textwidth]{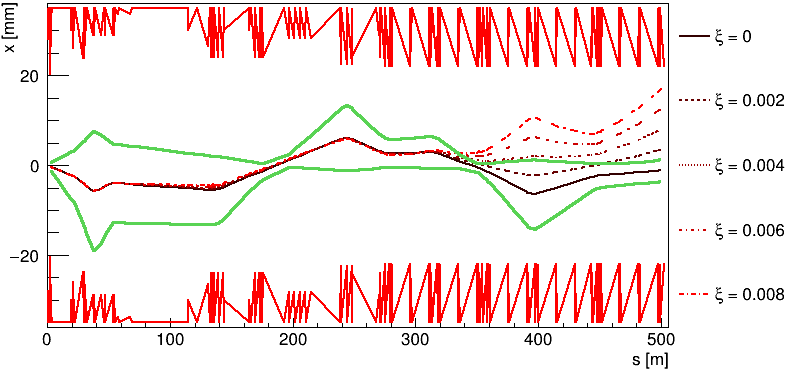}}
\caption{Energy dependence of the proton trajectory in $(x, s)$ plane for beam 1 around IP8 $(0,0)$. Protons were generated with the transverse momentum set accordingly to -115 $\mu$rad value of the crossing angle. For detailed description of the elements see text.}
\label{fig_prot_traj_2}
\end{figure}

As indicated in Ref. \cite{FP420}, installing proton detectors more than 400 meters away from the IP, should give an access to much lower $\xi$ values. This is shown in Fig. \ref{fig_prot_traj_2}, where trajectories of protons which lost 2, 4, 6 and 8 $\permil$ are plotted. For the clarity magnets are not drawn. From studies of optics files one can conclude that a possible place to install a Hamburg beampipe is around 430 m.

\section{Geometric Acceptance}
\begin{figure}[htb]
\centerline{%
  \includegraphics[width=0.33\textwidth]{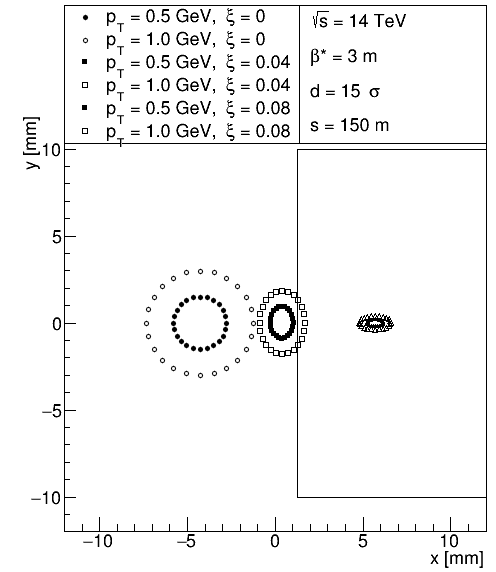}
  \includegraphics[width=0.33\textwidth]{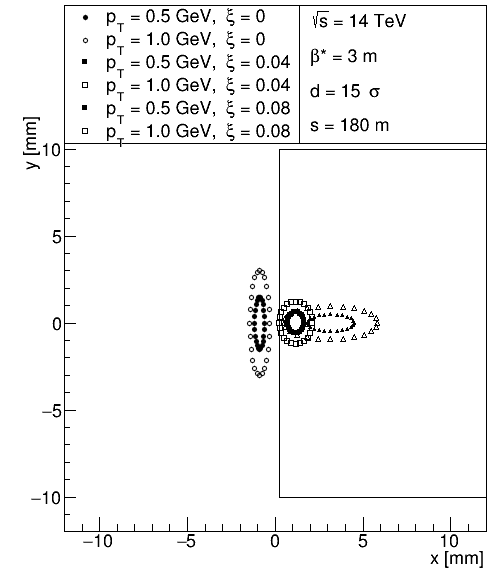}
  \includegraphics[width=0.33\textwidth]{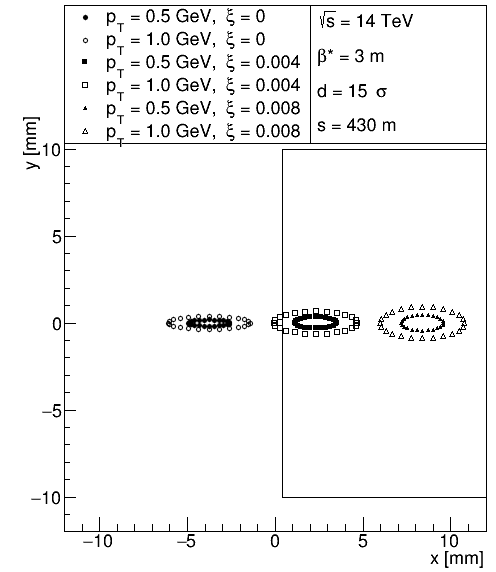}
}
\caption{Proton positions with different relative energy loss ($\xi$) and transverse momentum ($p_T$) for three considered detector locations. The solid lines mark the possible proton detector. The distance from the beam center was set to $15\sigma$.}
\label{fig_pos}
\end{figure}

For the feasibility studies it is important to understand the connection between the kinematics of scattered proton and its position at the detector location. These dependences, for three possible detector locations (150, 180 and 430 m from IP8), are shown in Fig. \ref{fig_pos}. To guide the eye, the solid lines indicate a possible detector shape.

\begin{figure}[htb]
\centerline{%
  \includegraphics[width=0.33\textwidth]{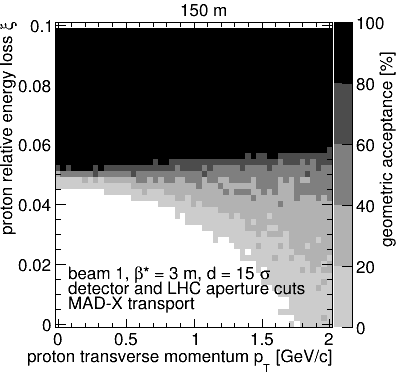}
  \includegraphics[width=0.33\textwidth]{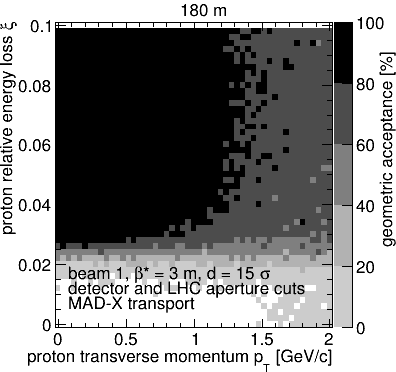}
  \includegraphics[width=0.33\textwidth]{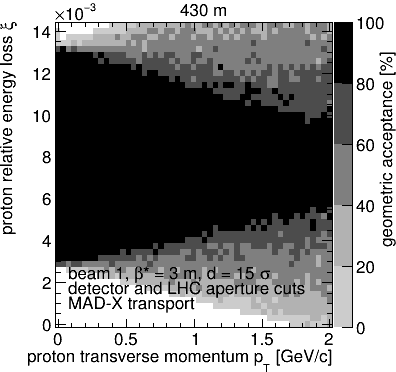}
}
\caption{Geometrical acceptance for three possible detector locations. Considered limitations: shape of the beam chamber, detector geometry and detector-beam distance. Please note a different vertical scale on the right plot.}
\label{fig_acc}
\end{figure}

Another useful information comes from the geometric acceptance defined as the ratio of the number of protons with a given relative energy loss and transverse momentum ($p_T$) that reached the detector to the total number of the scattered protons having $\xi$ and $p_T$. Acceptances for all considered cases are shown in Fig. \ref{fig_acc}. It should be noted that collimators were not considered in this studies and the upper limit of acceptance (10\%) comes from the internal settings of MAD-X PTC\footnote{This should not be a major drawback since the experience of ATLAS and CMS/TOTEM Collaborations shows that during the regular LHC runs collimators absorb protons having $\xi \gtrsim 0.1$.}. Nevertheless, it can be concluded that placing detectors at around 150 m and 180 m from IP8 will result in $\xi$ acceptance above 0.05 and 0.025, respectively. Hamburg beampipe placed about 430 m from IP8 should allow access to protons with $0.003 < \xi < 0.013$.

\section{Mass Acceptance}
Energy of scattered protons is precisely correlated with the energy of central system: $\xi_{1,2} = M_X \cdot \exp(\pm y) /\sqrt{s}$, where $M_X$ ($y$) is mass (rapidity) of the central system and $\xi_{1,2}$ denote proton scattered into beam 1 and 2, correspondingly.

\begin{figure}[htb]
\centerline{%
  \includegraphics[width=0.49\textwidth]{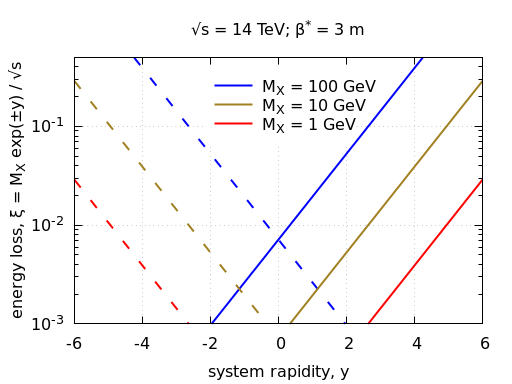}
  \includegraphics[width=0.49\textwidth]{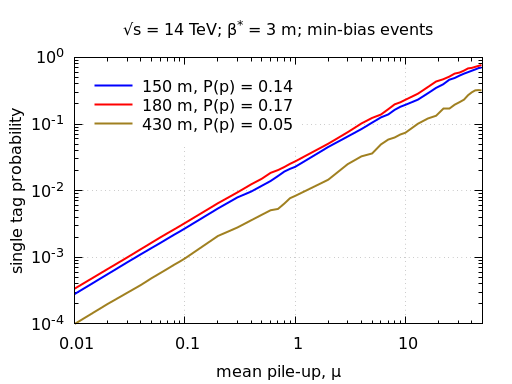}
}
\caption{\textbf{Left:} relation between proton energy loss ($\xi$) and rapidity ($y$) at which central system ($M_X$) is produced; solid (dashed) lines represent $\xi$ in the beam 1 (2) direction. \textbf{Right:} single tag probability as a function of mean pile-up; probabilities of having a minimum bias proton, P(p), in a given detector location were obtained by Pythia 8. Detector-beam distance was assumed to be 15$\sigma$.}
\label{fig_mass_mu}
\end{figure}

In case of diffractive production a part of energy is carried by proton or Pomeron remnants. On the contrary, in case of the exclusive events all energy is used to produce central system. The asymmetric acceptance of the LHCb detector ($2 < \eta < 5$) results in asymmetry between $\xi_1$ and $\xi_2$. \textit{I.e.} the boost towards positive $y$ should result in $\xi_1$ higher than $\xi_2$. Such feature will have a major impact on the possible measurement programme. This is shown in Fig. \ref{fig_mass_mu} (left). For example, the exclusive $J/\Psi$ production \cite{EXC_JPSI} measurement would be possible with a single proton tag assuming detector at 430 m (beam 1), whereas the exclusive Higgs production \cite{EXC_Higgs} with a single tag should be measurable with detectors placed around 150 -- 180 m (beam 1). Double tagged exclusive Higgs measurement might be possible with additional set of forward proton detectors positioned around 430 m on beam~2.

\section{Impact of Pile-up}
As was shown in \textit{e.g.} \cite{LHC_forward_physics, EXC_DT, EXC_ST1, EXC_ST2}, the main background for the diffractive and exclusive events at the LHC comes from a non-diffractive production overlaid with pile-up protons. This is presumably true also for LHCb. Probabilities of having a proton tag within the acceptance of considered forward proton detectors are shown in Fig. \ref{fig_mass_mu} (right). Probabilities of having a minimum bias proton in a given detector location were obtained by Pythia~8 \cite{Pythia8}. As can be concluded from the figure, the possibility of proton tagging provides a non-diffractive background reduction by a factor of 10 -- 30 for $\mu \sim 5$ depending on the detector placement. However, one should note that the beam-induced background (\textit{i.a.} halo) which will be an important ingredient especially at low-$\xi$ was not considered.

\section{Outlook}
Initial studies presented in this paper indicate the possibility of having forward proton detectors around LHCb starting from Run 4. Presented acceptances are one of the first steps for dedicated feasibility studies, based on which the physics programme should be composed. Moderate values of pile-up give hope of a interesting measurements complementary to the ones done by ATLAS and CMS/TOTEM Collaborations.



\end{document}